%% file: main.tex
\documentclass[10pt,conference]{IEEEtran}
\IEEEoverridecommandlockouts

\usepackage{silence}
\usepackage{cite}

\usepackage{amsmath,amssymb,amsthm,fixmath}

\usepackage{color,colortbl}
\usepackage{xcolor}

\usepackage[inline]{enumitem}

\usepackage{siunitx}
\DeclareSIUnit{\dBm}{dBm}

\usepackage{graphicx}
\usepackage[labelformat=simple]{subcaption}

\usepackage{epstopdf}
\usepackage{multirow,booktabs}
\usepackage{diagbox}
\usepackage{makecell}
\usepackage{hhline}
\usepackage{array} 
\newcolumntype{x}{!{\vrule width 2px}}
\newcolumntype{y}{!{\vrule width 1.5px}}

\usepackage{optidef}

\usepackage[shortcuts,acronym,automake]{glossaries}
\input{glossary}

\usepackage{tcolorbox}
\tcbuselibrary{many}

\usepackage[norelsize,linesnumbered,ruled]{algorithm2e}
\SetKwRepeat{Do}{do}{while}
\makeatletter
\newcommand{\removelatexerror} {\let\@latex@error\@gobble}
\makeatother

\usepackage[normalem]{ulem}

\usepackage[textsize=tiny,colorinlistoftodos]{todonotes}

\tikzstyle{note}=[rectangle, minimum width=3cm, draw = none, fill = none, minimum width = 1.5cm, anchor=center, align=left]
\tikzstyle{block}=[rectangle, draw, line width=1pt, fill = none, minimum width = 1cm, minimum height = 0.75cm, anchor=center, inner sep = 0.5mm, align=center]
\tikzstyle{arrow} = [thick,->,>=stealth]

\newif\ifreviewmode
\reviewmodetrue 

\ifreviewmode
\else
  \renewcommand{\todo}[1]{} 
\fi

\begin{document}

\title{Two-Timescale Dynamic Service Deployment and Task Scheduling with Spatiotemporal Collaboration in Mobile Edge Networks}

\author{
	\IEEEauthorblockN{
		Yang~Li\IEEEauthorrefmark{2},
		Xing~Zhang\IEEEauthorrefmark{2}\IEEEauthorrefmark{1},
		Yunji~Zhao\IEEEauthorrefmark{2},
		and~Wenbo~Wang\IEEEauthorrefmark{2}
	}
	\IEEEauthorblockA{
		\IEEEauthorrefmark{2}Key Laboratory of Universal Wireless Communications, Ministry of Education\\
		\IEEEauthorrefmark{2}Beijing University of Posts and Telecommunications, Beijing 100876, China\\
            \IEEEauthorrefmark{1}Email: zhangx@ieee.org
	}
}

\maketitle

\begin{abstract}
Collaborative edge computing addresses the resource constraints of individual edge nodes by enabling resource sharing and task co-processing across multiple nodes. To fully leverage the advantages of collaborative edge computing, joint optimization of service deployment and task scheduling is necessary. Existing optimization methods insufficiently address the collaboration across spatial and temporal dimensions, which hinders their adaptability to the spatiotemporally varying nature of user demands and system states. This paper focuses on optimizing the expected task processing delay in edge networks. We propose a two-timescale online optimization framework to jointly determine: \textit{i)} service deployment decisions at each large timescale; and \textit{ii)} task scheduling decisions at each small timescale. Specifically, the convex optimization technique is used to solve the task scheduling problem, while a multi-agent deep reinforcement learning technique is employed for the service deployment problem. These two methods are combined for spatiotemporal co-optimization through a two-timescale alternating optimization approach. Compared to the baseline algorithms, the proposed scheme achieves better delay performance, while also exhibiting low running time and favorable convergence behavior.
\end{abstract}

\section{Introduction}\label{sec:I}

Mobile edge computing (MEC) extends cloud computing and storage resources to the network edge, such as base stations (BSs), to deliver low-latency and energy-efficient services to nearby users \cite{1}. Compared with mobile cloud computing, which provides elastic resource capacities, MEC faces significant limitations due to the constrained and heterogeneous resources of edge nodes. To address these limitations, collaborative edge computing (CEC) schemes have been proposed and have attracted considerable attention \cite{2}. Service deployment \cite{4} and task scheduling \cite{3} are fundamental challenges in CEC. However, the interdependence between service deployment and task scheduling, combined with the spatio-temporal non-uniformity of user demands, greatly increases the complexity of their joint optimization, making it a hot topic in recent years’ research.

Several recent studies have employed online optimization methods to jointly optimize service deployment and task scheduling in collaborative edge networks. For example, the authors in \cite{5} employed a deep reinforcement learning (DRL)-based algorithm to jointly optimize UAV trajectory planning, task scheduling, and service function deployment in a UAV-enabled edge computing scenario. In \cite{6}, a dynamic service deployment and task scheduling strategy was proposed for a dual-service pooling-based hierarchical cloud service system in intelligent buildings. Specifically, an analytic hierarchy process (AHP)-based quality of service (QoS) evaluation mechanism and a dynamic normal distribution selection method were applied for service deployment, while a dynamic inertia particle swarm optimization (DI-PSO) algorithm was adopted for task scheduling. In \cite{7}, the unpredictable spatiotemporal patterns of service requests in MEC systems were considered, and an online optimization framework was designed to enhance long-term system performance through collaborative service placement, task scheduling, and resource allocation. 

Furthermore, distinguishing from the single-timescale decision-making in the above studies, some preliminary studies optimize service deployment and task scheduling asynchronously across different timescales. The authors in \cite{8} proposed a two-timescale online optimization algorithm that jointly optimizes microservice container placement and task scheduling to enhance microservice deployment reliability in MEC. In \cite{9}, a two-timescale optimization approach was proposed to address task offloading, resource allocation, and model deployment for AI as a Service (AIaaS) in future 6G networks. In \cite{10}, the joint service deployment and task scheduling problem in satellite edge computing was investigated. A two-timescale hierarchical optimization framework was proposed, significantly improving service performance while reducing deployment costs and energy consumption. 

Although existing studies have considered the coupling between service deployment and task scheduling, as well as the time-varying nature of system states, they have rarely addressed the joint optimization of both in temporal and spatial dimensions. In edge computing scenarios, spatiotemporal features are particularly prominent, making spatiotemporal coordination optimization crucial for enhancing system performance. To overcome this challenge, this paper proposes a novel two-timescale online optimization framework for spatiotemporal co-enhancement, comprising: \textit{i)} making service deployment decisions for each ES at the large timescale, and \textit{ii)} determining task scheduling strategies for each service at the small timescale. The main contributions of this paper are outlined as follows:
\begin{itemize}
\item{To minimize the expected processing delay of user requests in edge networks, a mixed-integer nonlinear programming (MINLP) problem is proposed, which is NP-hard.}
\item{A spatiotemporal collaboration-enhanced two-timescale online optimization scheme is introduced, integrating convex optimization with multi-agent DRL techniques.}
\item{Extensive experiments show that the proposed scheme outperforms other methods in terms of delay performance and demonstrates low running time and favorable convergence behavior.}
\end{itemize}

\begin{figure}[h]
\centerline{\includegraphics[width=0.45\textwidth]{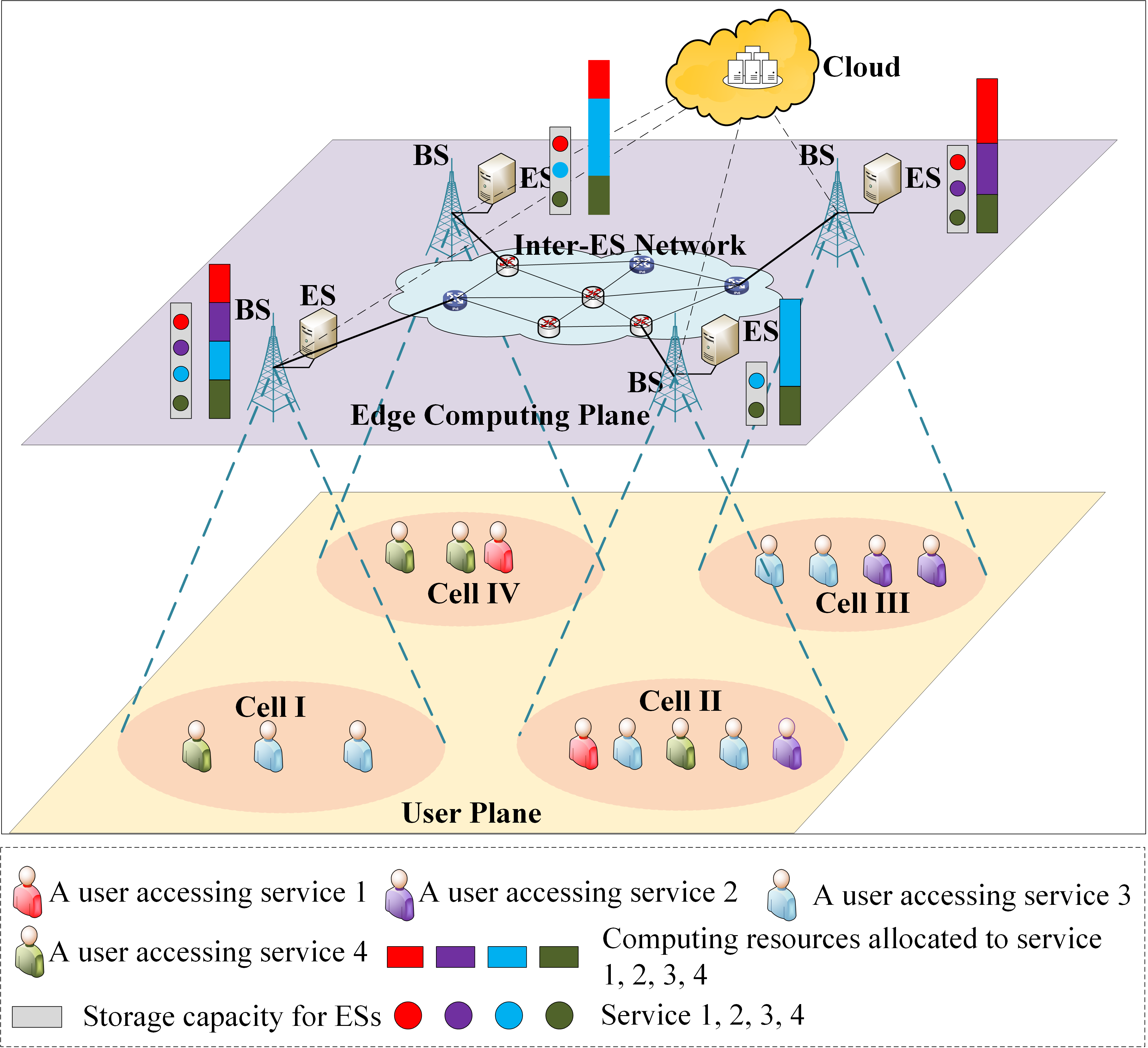}}
    \caption{An illustration of the considered system.}
\label{system}
\vspace{-0.5cm}
\end{figure}

\section{System Model}\label{sec:II}
\subsection{Overview of the System}\label{sec:II-A}
As illustrated in Fig. \ref{system}, we consider an MEC system composed of multiple BSs and ESs, both indexed by the set $\mathcal{M} = \{1, 2, ..., M\}$, and geographically distributed across various locations. The system also includes a remote cloud. Multiple users exist, each associated with a BS within the same cell. Each ES can offer computing services to users, provided the corresponding applications (services) are pre-deployed on it. All services are represented by the set $\mathcal{J} = \{1, 2, \dots, J\}$. Since ESs are resource-constrained compared to the cloud, each can deploy only a subset of services, while the cloud can deploy the entire set. When a task arrives, if the ES has not deployed the corresponding service, it can only be scheduled to nearby available edge nodes or processed on the cloud. In practice, user request distribution exhibits spatiotemporal non-uniformity, necessitating dynamic adjustment of task scheduling and service deployment to optimize users’ quality of service (QoS).

\begin{figure}[h]
\centerline{\includegraphics[width=0.47\textwidth]{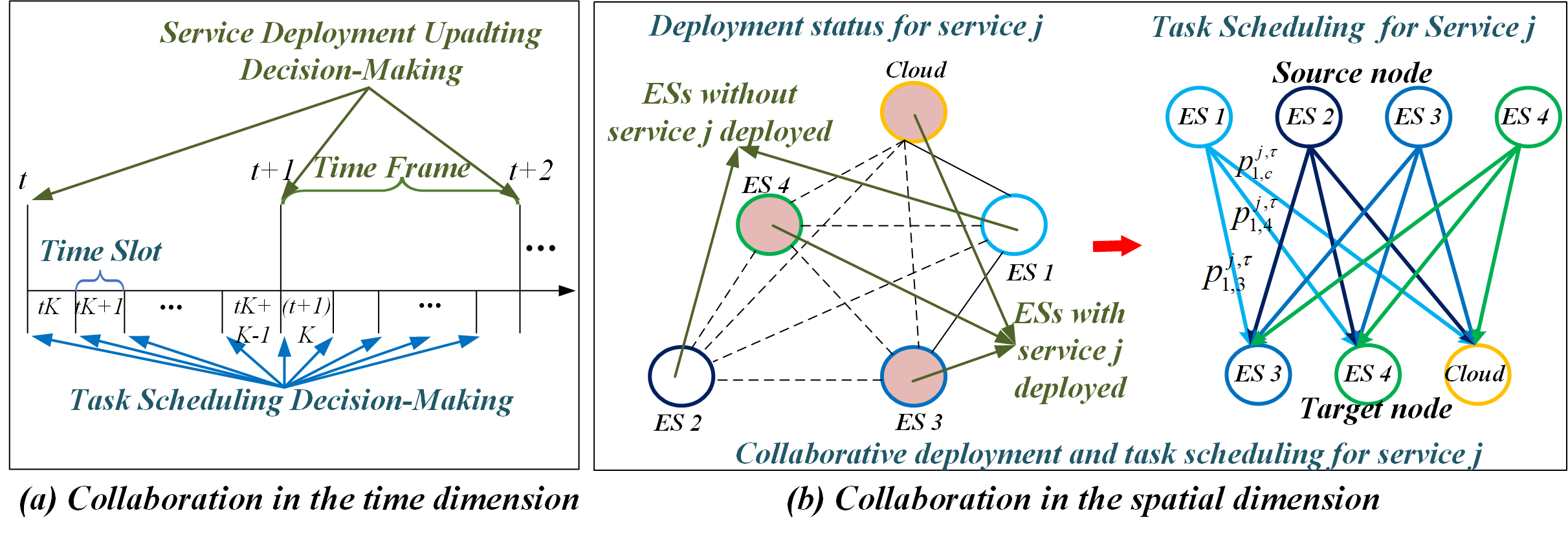}}
    \caption{Spatiotemporal collaboration in the two-timescale online optimization framework.}
\label{timespace}
\vspace{-0.5cm}
\end{figure}

Additionally, task scheduling and service deployment decisions have different time sensitivities. Making them simultaneously would cause the task scheduling process to wait for service deployment, reducing its real-time requirements \cite{10}. Therefore, we design an online optimization framework where service deployment decisions are made at each large time scale, while task scheduling decisions are made at each smaller time scale, as shown in Fig. \ref{timespace}(a). Specifically, time is divided into $T \in \mathbb{N}^+$ coarse-grained time frames, each of which can be further divided into $K \in \mathbb{N}^+$ fine-grained time slots. $t\in \{0,1,\dots,T-1\}$ represents the index of the $t$-th time frame, and $\tau \in \mathcal{T}_t = \{tK, tK+1, \dots, tK+K-1\}$ is the index of the $\tau$-th time slot in the $t$-th time frame.

\subsection{Service Deployment Model}\label{sec:II-B}
Let $\mathcal{D}_m^t = \{d_m^{1,t}, d_m^{2,t}, \dots, d_m^{J,t}\}$ denote the service deployment decision of ES $m$ at the $t$-th time frame. Here, $d_m^{j,t} = 1$ indicates that service $j$ is deployed on ES $m$, while $d_m^{j,t} = 0$ means that service $j$ is not deployed. Given the limited storage capacity of ES $m$, its service deployment decisions at each time frame must satisfy the following condition:
\begin{align}
	\sum_{j\in\mathcal{J}}d_m^{j,t} V_j\le C_m,\label{eq:storage_condition}
\end{align}
where $V_j$ denotes the data size of service $j$, and $C_m$ represents the storage capacity of ES $m$. Following service deployment, the corresponding computation and bandwidth resources must be allocated for each service. Let $f_m^{j,t}$ denote the computation resources allocated to service $j$ by ES $m$ in the $t$-th time frame, and $R_{m,m^{\prime}}^{j,t}$ represents the bandwidth allocated to the task scheduling for service $j$ from ES $m$ to ES $m^{\prime}$ in the $t$-th time frame. Clearly, $R_{m,m}^{j,t} = \infty$. The specific values of these variables depend on the service deployment decisions. Since this paper focuses on the joint optimization of service deployment and task scheduling, we model these variables as functions of the service deployment decisions. The specific optimization of resource allocation strategies will be addressed in future research.
\subsection{Task Scheduling Model}\label{sec:II-C}
A task can only be processed on the computation node where its corresponding service is deployed. As shown in Fig. 2(b), we define the task scheduling decision of ES $m \in \mathcal{M}$ for service $j$ during time slot $\tau \in \mathcal{T}_t$ as $\mathcal{P}_m^{j,\tau}=\{p_{m, n}^{j,\tau}|n\in \mathcal{M}\cup\{c\}\}$, where $c$ represents the cloud. In this context, $p_{m,n}^{j,\tau}$ represents the probability that ES $m$ schedules a task for service $j$ to computation node $n$ for processing. The task scheduling decision of ES $m$ for service $j$ must satisfy the following conditions:
\vspace{-0.1cm}
\begin{align}
	 &p_{m,m^{\prime}}^{j,\tau}=0\quad if\; d_{m^{\prime}}^{j,t}=0,\forall m^{\prime} \in \mathcal{M}\label{eq:scheduling_condition1} \\
	&\sum_{n \in \mathcal{M}\cup \{c\}}p_{m,n}^{j,\tau} =1.\label{eq:scheduling_condition2}
\end{align}
\subsection{Processing Delay Model}\label{sec:II-D}
In this scenario, the processing delay of a task in the edge network is influenced by both scheduling and computation delays. The wireless transmission delay for task offloading and the result return delay are excluded, as service deployment and task scheduling decisions do not impact the former, and the latter is negligible due to the small size of the result.

We begin by analyzing the expected processing delay for a user-initiated task for service $j$ arriving at ES $m$ at the beginning of time slot $\tau$. Assume that all tasks for a service processed on a ES during a time slot equally share the computational resources allocated to that service. Once the task is processed on ES $m^{\prime}$, its expected processing delay can be calculated as
\begin{align}
	T_{m,m^{\prime}}^{j,\tau} = \frac{S_j}{R_{m,m^{\prime}}^{j,t}}+\frac{ \lambda_j}{f_{m^{\prime}}^{j,\tau}},\label{eq:exp_delay1}
\end{align}
where $S_j$ denotes the size of a task for service $j$, and $\lambda_j$ represents the number of CPU cycles required for processing it. According to the previous assumption, the expected amount of computational resources allocated to a task for service $j$ on ES $m^{\prime}$ in time slot $\tau \in \mathcal{T}_{\tau}$ is given by $f_{m^{\prime}}^{j,\tau} = \frac{f_{m^{\prime}}^{j,t}}{\sum_{m \in \mathcal{M}} p_{m,m^{\prime}}^{j,\tau} N_m^{j,\tau}}$. Here, $N_m^{j,\tau}$ denotes the number of user-initiated tasks for service $j$ that arrive at ES $m$ at the beginning of time slot $\tau$.

If the task is processed on the cloud, its expected processing delay can be calculated as
\vspace{-0.1cm}
\begin{align}
	T_{m,c}^{j,\tau} = \frac{S_j}{R_{m,c}^t}+\frac{ \lambda_j}{f_{c}},\label{eq:exp_delay2}
\end{align}
where $R_{m,c}^t$ denotes the transmission rate between ES $m$ and the cloud during the $t$-th time frame. Due to the longer backhaul link, its value is lower than the transmission rates between ESs. $f_c$ denotes the computational resources allocated by the cloud to each task. Given the ample computational resources available in the cloud, $f_c$ is assumed to be a constant greater than $f_m^{j,\tau}, \forall m\in \mathcal{M}$.

Based on the above analysis, the expected processing delay of a task for service $j$ that arrives at ES $m$ at the beginning of time slot $\tau$ can be calculated as 
\begin{align}
	T_{m}^{j,\tau} = \sum_{m^{\prime}\in \mathcal{M}}p_{m,m^{\prime}}^{j,\tau}T_{m,m^{\prime}}^{j,\tau}+p_{m,c}^{j,\tau} T_{m,c}^{j,\tau} .\label{eq:exp_delay3}
\end{align}

Accordingly, the expected processing delay of a task for service $j$ in the edge network can be expressed as
\begin{align}
	T^{j,\tau} = \sum_{m\in \mathcal{M}}\frac{N_m^{j,\tau}}{\sum_{m^{\prime}\in \mathcal{M}}N_{m^{\prime}}^{j,\tau}}T_{m}^{j,\tau}.\label{eq:exp_delay4}
\end{align}
Finally, the expected processing delay for any task during time slot $\tau$ can be expressed as
\begin{align}
	T^\tau =\sum_{j\in \mathcal{J}}\frac{N^{j,\tau}}{\sum_{j^{\prime}\in \mathcal{J}}N^{j^{\prime},\tau}} T^{j,\tau},\label{eq:exp_delay5}
\end{align}
where $N^{j,\tau} = \sum_{m \in \mathcal{M}} N_m^{j,\tau}$ denotes the total number of tasks for service $j$ received by the entire edge network at the beginning of time slot $\tau$.

\section{Problem Formulation}\label{sec:problem}
To enhance the QoS for all users in the considered MEC system, our objective is to minimize the expected processing delay of tasks in the edge network. Given the time-varying nature of the MEC system, it is necessary to optimize the relevant decisions from a long-term and dynamic perspective. Accordingly, the time-average expected processing delay of tasks in the edge network is adopted as the performance metric and is given by
\vspace{-0.5cm}
\begin{align}
	\mathcal{F}_T = \frac{1}{KT}\sum_{t=1}^T\sum_{\tau \in \mathcal{T}_t}T^{\tau}.\label{eq:objective}
\end{align}

To minimize $\mathcal{F}_T$, we jointly optimize two sets of decisions across different timescales: \textit{(i)} the service deployment decisions of each ES in each time frame, and \textit{(ii)} the task scheduling decisions of each ES in each time slot. Let $\mathcal{V}_A^t=\{\mathcal{D}_{m}^t|m\in \mathcal{M}\}$, $\mathcal{V}_B^{\tau}=\{\mathcal{P}_m^{j,\tau}|m\in \mathcal{M},j\in\mathcal{J}\}$, the problem can be formulated as
\begin{align}
\mathcal{P}_1: \quad &min_{\mathcal{V}_A^t,\mathcal{V}_B^{\tau}} lim_{T \rightarrow \infty} \mathcal{F}_T \\
\text{s.t.} \quad &(10a):(\ref{eq:storage_condition}),(\ref{eq:scheduling_condition1}),(\ref{eq:scheduling_condition2}) \notag \\
&(10b):d_m^{j,t} \in \{0,1\},\quad \forall m\in \mathcal{M},\forall j\in \mathcal{J} \notag \\
&(10c):p_{m,n}^{j,\tau} \in [0,1],\quad  \forall n\in \mathcal{M}\cup \{c\}, \forall m,j \notag 
\end{align}

Constraints (10b) and (10c) define the feasible domains of the service deployment and task scheduling decision variables, respectively. Clearly, the problem $\mathcal{P}_1$ is a two-timescale MINLP problem, which is NP-hard \cite{12}.
\section{Two-Timescale Optimization Algorithm}\label{sec:algorithm}
In this section, we propose a spatiotemporal collaboration-enhanced two-timescale optimization approach to jointly optimize service deployment and task scheduling.
\subsection{Problem Decomposition}\label{subsec:decoupling}
To solve problem $\mathcal{P}_1$, we first decouple the original problem into two subproblems corresponding to different time scales. Since the service deployment status remains fixed within a time frame, the task scheduling problem can be decomposed into independent subproblems for each time slot. For a given time slot $\tau \in \mathcal{T}_t$, the corresponding task scheduling subproblem can be formulated as follows:
\begin{align}
\mathcal{P}_2: \quad &min_{\mathcal{V}_B^{\tau}} T^{\tau} \\
\text{s.t.} \quad &(11a):(\ref{eq:scheduling_condition1}),(\ref{eq:scheduling_condition2}),(10c)\notag 
\end{align}

In $\mathcal{P}_2$, the service deployment state is given. Then, we can solve the service deployment problem formulated as follows:
\begin{align}
\mathcal{P}_3: \quad &min_{\mathcal{V}_A^t} lim_{T \rightarrow \infty} \mathcal{F}_T \\
\text{s.t.} \quad &(12a):(\ref{eq:storage_condition}),(10b)\notag \\
&(12b):p_{m,n}^{j,\tau} = p_{m,n}^{j,\tau *}\quad\forall n\in \mathcal{M}\cup \{c\}, \forall \tau ,m,j, \notag 
\end{align}
where $p_{m,n}^{j,\tau *}$ represents the optimal solution of $\mathcal{P}_2$. This implies that, although $\mathcal{P}_3$ targets optimization over large time scales, it still depends on decision variables defined over small time scales.

\subsection{Small-Timescale Decisions: Task Scheduling Optimization}\label{subsec:Task Scheduling}
For problem $\mathcal{P}_2$, the task scheduling of each service is observed to be mutually independent. Therefore, $\mathcal{P}_2$ can be decomposed into $J$ independent subproblems. The $j$-th subproblem focuses on optimizing the task scheduling for service $j$ in time slot $\tau$, and is formulated as follows:
\begin{align}
\mathcal{P}_{4-j}: \quad &min_{\mathcal{V}_B^{j,\tau}}T^{j,\tau} \\
\text{s.t.} \quad &(13a):(\ref{eq:scheduling_condition1}),(\ref{eq:scheduling_condition2})\notag \\
&(13b):p_{m,n}^{j,\tau} \in [0,1],\quad \forall n\in \mathcal{M}\cup \{c\},  \forall m \notag 
\end{align}

where $\mathcal{V}_B^{j,\tau} = \{\mathcal{P}_m^{j,\tau} \mid m \in \mathcal{M}\}$. The constraints are linear with respect to the optimization variables, while the objective function is a nonlinear convex function, which can be verified through its second-order derivatives. Therefore, the problem is classified as a convex optimization problem. A variety of low-complexity methods can be employed to solve this problem, and detailed discussions are omitted here. In our simulations, the problem is solved using the CVXPY (a software-based optimization solver) \cite{12}.

\subsection{Large-Timescale Decisions: Service Deployment Optimization}\label{subsec:Service Deployment}
Intuitively, the service deployment decisions made at each time frame directly determine the upper bound of the performance that task scheduling can achieve within the corresponding time slots. Due to the spatiotemporal non-uniformity of user request distributions, the service deployment strategy need to enable dynamic spatiotemporal collaborative optimization. By adapting to temporal and spatial variations, this approach enhances resource utilization and task scheduling efficiency. In addition, as illustrated in Fig. 2, the two-timescale dynamic service deployment and task scheduling inherently exhibit spatiotemporal coupling, which should be fully exploited.

\begin{figure}[t]
\centerline{\includegraphics[width=0.45\textwidth]{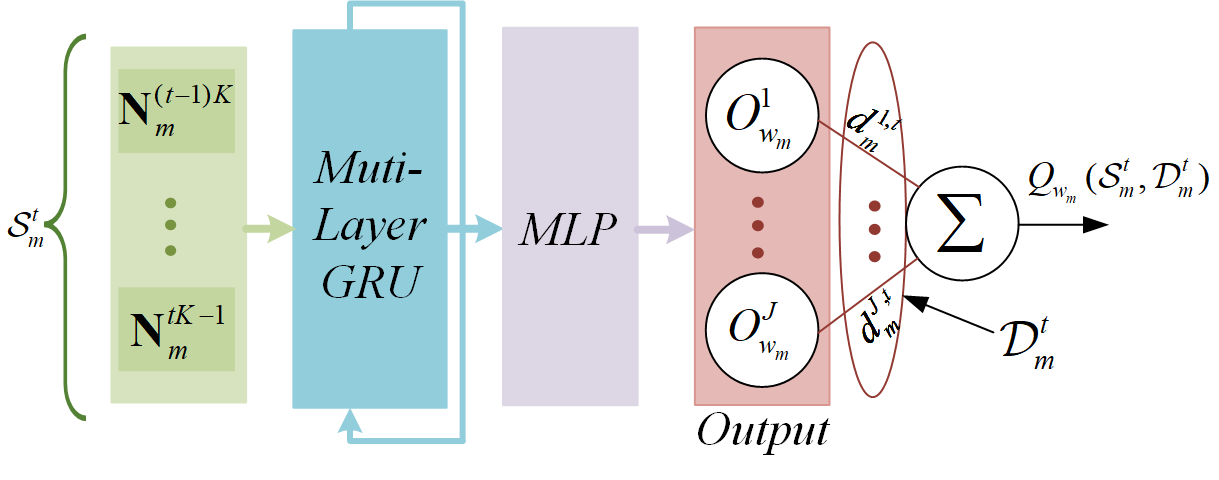}}
    \caption{The architecture of the agent network for ES $m$.}
\label{architecture}
\vspace{-0.5cm}
\end{figure}
Based on the above analysis, for problem $\mathcal{P}_3$, a multi-agent DRL algorithm is adopted to jointly optimize the service deployment decisions across different ESs. Specifically, $M$ agents are designed, where the $m$-th agent is deployed on ES $m$ and is responsible for generating the service deployment decision $\mathcal{D}_m^t$. To optimize each agent’s policy in an online manner, the double deep Q-network (DDQN) algorithm is employed. Therefore, each ES is equipped with both a training network and a target network that share the same architecture. The weights of the training and target networks on ES $m$ are denoted by $w_m$ and $w_m^-$, respectively. The observed state of the $m$-th agent at time frame $t$ is defined as the number of tasks for each service arriving at ES $m$ at the beginning of each time slot during the $(t-1)$-th frame, i.e., $\mathcal{S}_m^t = \{\boldsymbol{N}_m^{\tau} \mid \tau \in \mathcal{T}_{t-1}\}$, where $\boldsymbol{N}_m^{\tau} = [N_m^{1,\tau}, \dots, N_m^{J,\tau}]$. In addition, the action space $\mathcal{A}_m$ of the $m$-th agent comprises all feasible service deployment strategies satisfying constraint (\ref{eq:storage_condition}). To facilitate the spatiotemporal collaborative optimization of service deployment decisions made by different agents, the cumulative task scheduling performance gain over all time slots in time frame $t$ is used to evaluate the joint actions of all agents. This cumulative gain serves as the action reward $R^t$ for all agents. The task scheduling performance gain in time slot $\tau \in \mathcal{T}_t$ is defined as the reduction in the objective function value obtained by solving the problem $\mathcal{P}_2$ compared to the baseline scenario where all tasks are offloaded to the cloud.

To address the high-dimensional and complex action space introduced by numerous heterogeneous services, and to capture the temporal patterns of user request behavior, we draw inspiration from \cite{11} and design an enhanced model based on it. Specifically, for each ES, the intelligent agent network architecture is shown in Fig. \ref{architecture}. In the input layer, multiple gated recurrent unit (GRU) layers are employed to extract temporal features of user requests. A two-layer structure (TLS) is used as the output layer. The first layer of the TLS consists of $J$ neurons. The output of the $j$-th neuron, denoted as $O_{w_m}^j(\mathcal{S}_m^t)$, represents the partial state-action value associated with service $j$. The second layer of the TLS contains a single neuron without the activation unit. It outputs the predicted state-action value for the service deployment decision $\mathcal{D}_m^t$, denoted as $Q_{w_m}(\mathcal{S}_m^t,\mathcal{D}_m^t)$. This value is defined as the weighted sum of all partial state-action values, where the weights are determined by $\mathcal{D}_m^t$, i.e.,
\begin{align}
	Q_{w_m}(\mathcal{S}_m^t,\mathcal{D}_m^t) = \sum_{j\in \mathcal{J}}d_m^{j,t}O_{w_m}^j(\mathcal{S}_m^t).\label{eq:Q}
\end{align}

Notably, due to constraint (\ref{eq:storage_condition}), the optimal action for the $m$th intelligent agent must be determined by solving the following optimization problem:
\begin{align}
\mathcal{P}_{5}: \quad &max_{\mathcal{D}_m^t} \sum_{j\in \mathcal{J}}d_m^{j,t}O_{w_m}^j(\mathcal{S}_m^t) \\
\text{s.t.} \quad &(15a):(\ref{eq:storage_condition}),(10b)\notag
\end{align}
Problem $\mathcal{P}_5$ is a typical Knapsack problem, which can be efficiently solved using dynamic programming. For brevity, the detailed solution process is omitted.

\subsection{Proposed TTOSC Algorithm}\label{subsec:Algorithm}
This subsection introduces the two-timescale optimization method incorporating spatiotemporal collaboration (TTOSC), as illustrated in Algorithm~\ref{alg}.

Notably, to fully balance exploration and exploitation, the value of $\epsilon$ is annealed from an initial value to a small value close to zero over time.

\begin{algorithm}[h]
	\caption{Procedure of TTOSC }
	\label{alg}
	\small
	\DontPrintSemicolon
	For each ES, initialize an experience replay buffer with   
        capacity $E$, as well as set the target network update frequency $\delta$ and the discount factor $\gamma$.\; 
        For each ES, initialize the training network with random weights, and set the corresponding target network with the same initial weights.\;
	\For(){$t\in\{1,2,\dots T\}$}{
            Initialize $R^t=0$.\;
            For each ES, with probability $\epsilon$, its agent selects a stochastic service deployment action that satisfies constraint (1); with probability $1-\epsilon$, the agent determines the action by solving problem $\mathcal{P}_5$. The selected action is denoted by $A_m^t$. (When $t = 1$, since the system state is unobservable, the agent adopts a random policy.)\;

		\ForEach{$\tau\in\mathcal{T}_t$}{
                Solve problems $\mathcal{P}_{4-1}, \ldots, \mathcal{P}_{4-J}$ in parallel to obtain task scheduling decisions for all services, and calculate the objective value $T^{\tau}$ for time slot $\tau$.\;
                Calculate the objective value $\tilde{T}^{\tau}$ under the baseline scenario where all tasks are scheduled to the cloud for processing, and obtain the performance gain for time slot $\tau$ as $G^{\tau} = \tilde{T}^{\tau} - T^{\tau}$.\;
                Update $R^t=R^t+G^{\tau}$
		}
        
            If $t \ge 2$, for each ES, store the trajectory $(\mathcal{S}_m^{t}, A_m^{t}, R^{t}, \mathcal{S}_m^{t+1})$ into its corresponding experience replay buffer.\;
            For each ES, if there is enough data stored in its experience replay buffer, randomly sample $N$ trajectories $\{\mathcal{S}_{m,i}^{t},A_{m,i}^{t},R_i^{t},\mathcal{S}_{m,i}^{t+1}\}_{i=1,2,.... .N}$.\;
            For each ES and each sampled trajectory, the next state $\mathcal{S}_{m,i}^{t+1}$ is fed into the corresponding training network to compute partial state-action values. These values are subsequently utilized to solve problem $\mathcal{P}_5$, yielding the optimal service deployment action $A_{m,i}^{t+1*}$. Then, the target network is used to compute the expected state-action value (i.e., the temporal-difference target): $y_{m,i} = R_i^t + \gamma Q_{w_m^-}(S_{m,i}^{t+1}, A_{m,i}^{t+1*})$.\;
            For each ES, update the training network parameters by performing a gradient descent step on its loss function, which is defined as: $L_m=\frac{1}{N}\sum_{i=1}^N(y_{m,i}-Q_{w_m}(S_{m,i}^t,A_{m,i}^t))^2$.\;
            For each ES, the weight parameters of its training network are copied to the target network every $\delta$ time frames.\;
        
	}
\end{algorithm}

\section{Numerical Simulations}\label{sec:simualtion}
\subsection{Simulation Setup}\label{subsec:setup}
Consider an MEC system consisting of five cells by default, where the number of users in each cell is uniformly distributed between 10 and 50. A total of 20 services are provided in the system, and user requests for different services in each cell follow a Zipf distribution with an exponent of 1.2. Other network parameters are consistent with those in \cite{11}. In the agent network setup, the number of neurons in the hidden layer is set to 128, and the learning rate is 0.01. The initial value of $\epsilon$ is set to 0.5 and decays progressively to 0.01 over 500 episodes, after which it is held constant. During training, the batch size $N$ is 64, and the target network update frequency $\delta$ is 20. For simplicity in the simulation, we assume that each ES allocates computing and bandwidth resources equally to the deployed services. The following baseline schemes are used for performance comparison:
\begin{enumerate}
\item{\textit{\textbf{Greedy scheme}}: Service deployment and task scheduling decisions are made based on a greedy approach.}
\item{\textit{\textbf{DDPG-based scheme}}: The DDPG algorithm is used to select task scheduling policies at each time slot.}
\item{\textit{\textbf{Popularity-based scheme}}: Each ES deploys the services requested by users in the same cell during the previous time frame, prioritizing the most frequently requested services when its storage capacity is insufficient.}
\end{enumerate}

All the simulations are performed on a Pytorch 1.10.2 platform with an Intel Core i7-13650HX 4.9GHz CPU and 16 GB of memory.

\subsection{Simulation Results}
\begin{figure}[h]
\centering
\subfloat[]{\includegraphics[width=0.25\textwidth]{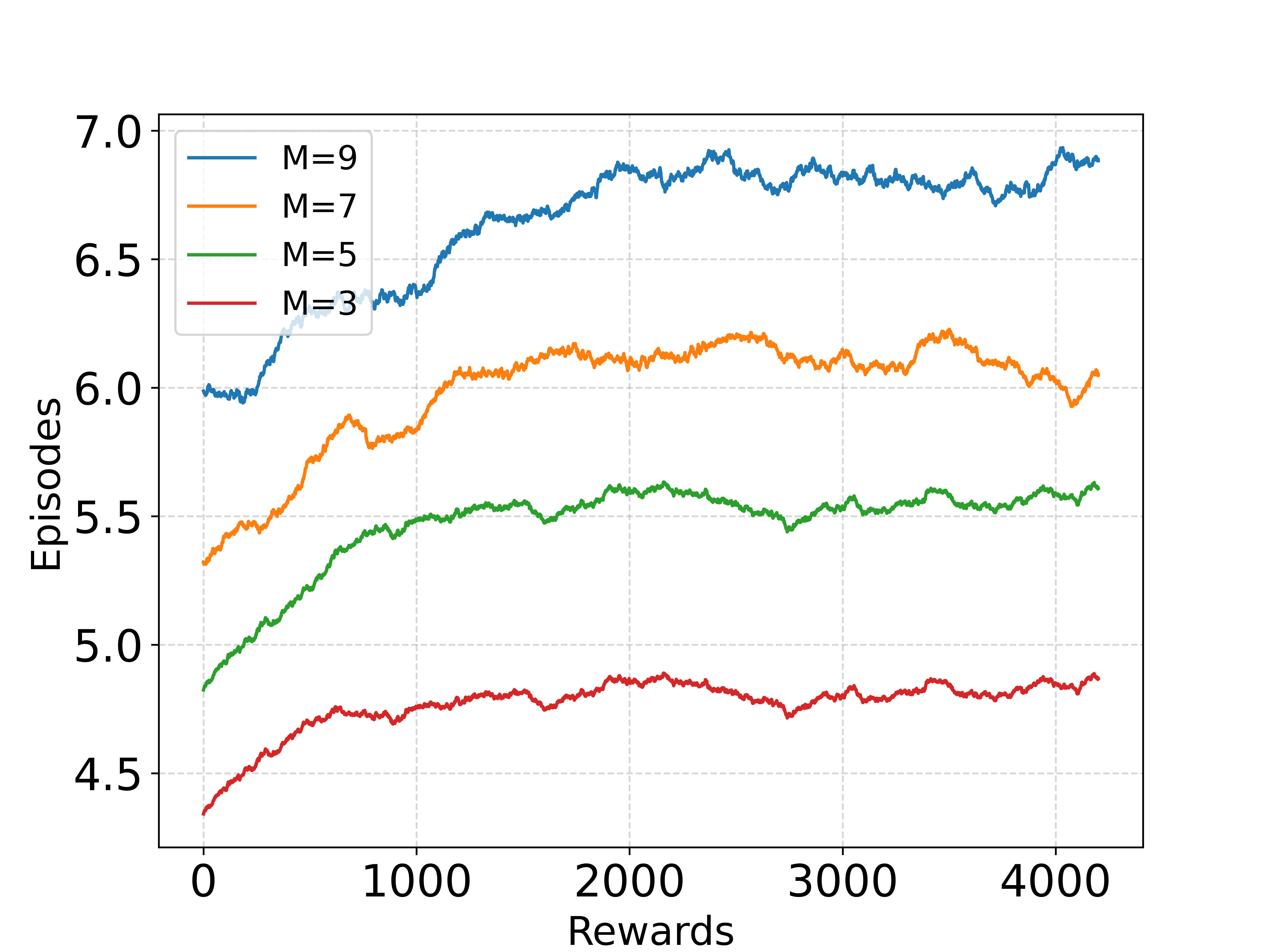}%
}
\subfloat[]{\includegraphics[width=0.25\textwidth]{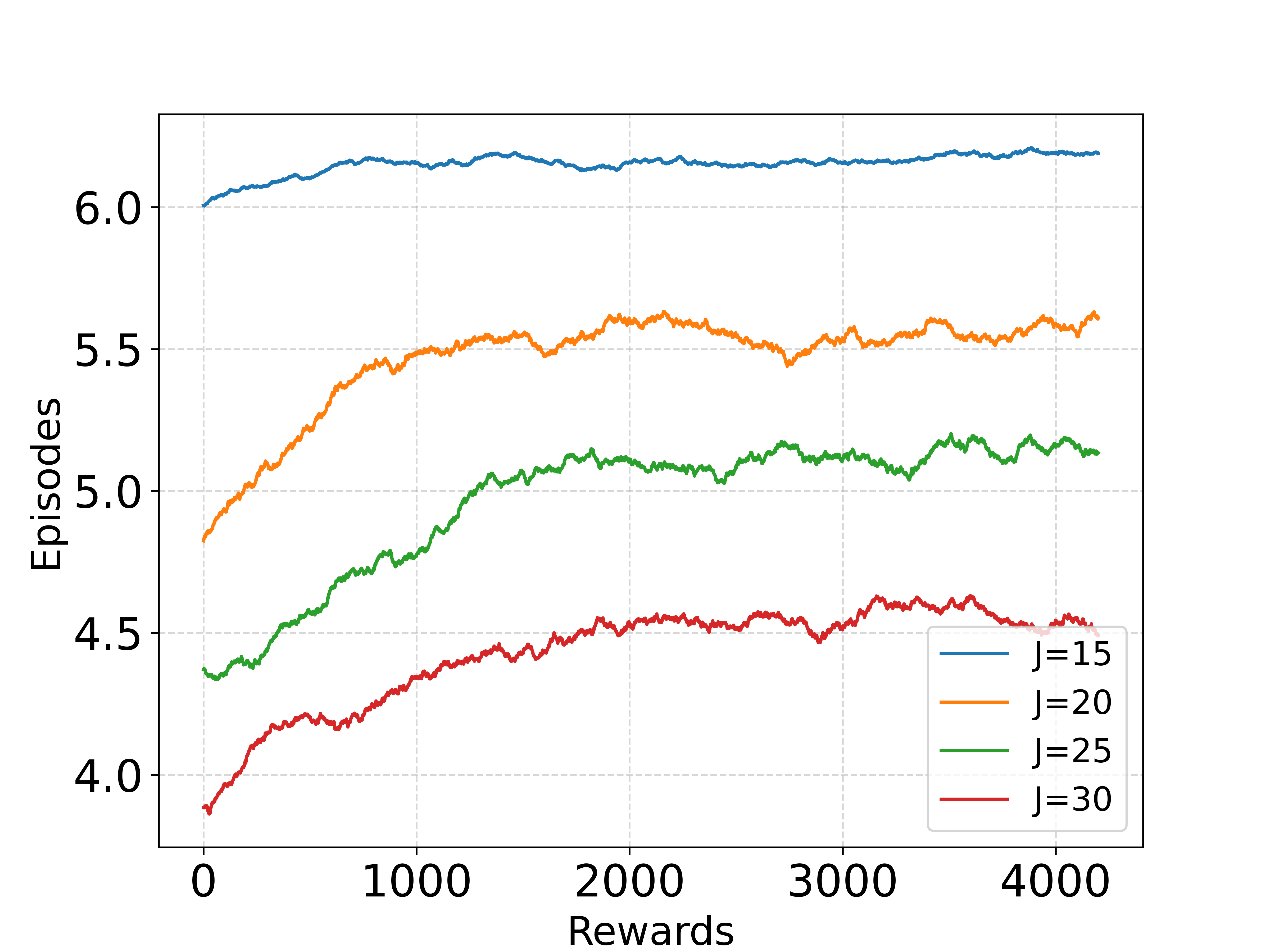}%
}
\vspace{-0.2cm}
\caption{Convergence performance of the proposed algorithm with (a) different numbers of cells and (b) different numbers of services.}
\label{fig4}
\end{figure}

As shown in Fig. \ref{fig4}, the proposed algorithm converges quickly. The learning curves fluctuate with each episode due to random factors in the generated data, such as uncertainties in task and edge network states. The curves in the figures are smoothed using the sliding window technique. As illustrated in Fig. \ref{fig4}(a), when $M = 3, 5, 7, 9$, approximately 800, 1000, 1500, and 2000 episodes are required for convergence, respectively. Therefore, the number of iterations required for convergence is expected to gradually increase as the number of cells grows. Additionally, we observe that as the number of cells increases, the average reward also grows.
\vspace{-0.2cm}

As shown in Fig. \ref{fig4}(b), as the number of services increases, the number of iterations required for convergence increases while the average reward decreases. This occurs because the limited storage capacity of ESs results in more requests needing to be scheduled to other available ESs or the cloud, which increases communication delays. Additionally, under the default configuration, the algorithm exhibits an average execution time of 0.00721 seconds for small-timescale task scheduling and 0.01939 seconds for large-timescale service deployment. These results demonstrate that the proposed algorithm enables near real-time decision-making.

\begin{figure}[h]
\centering
\subfloat[]{\includegraphics[width=0.24\textwidth]{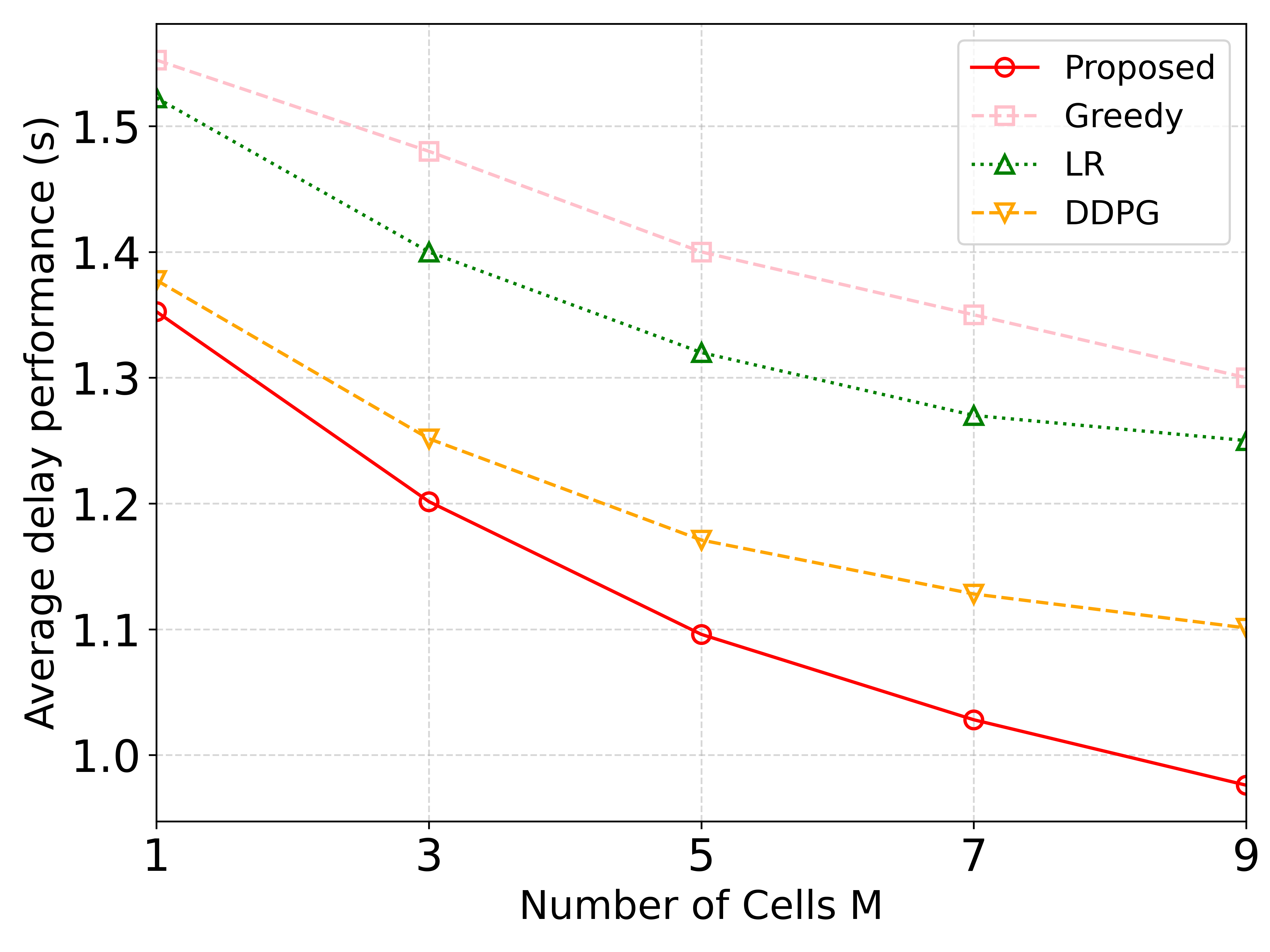}%
}
\subfloat[]{\includegraphics[width=0.24\textwidth]{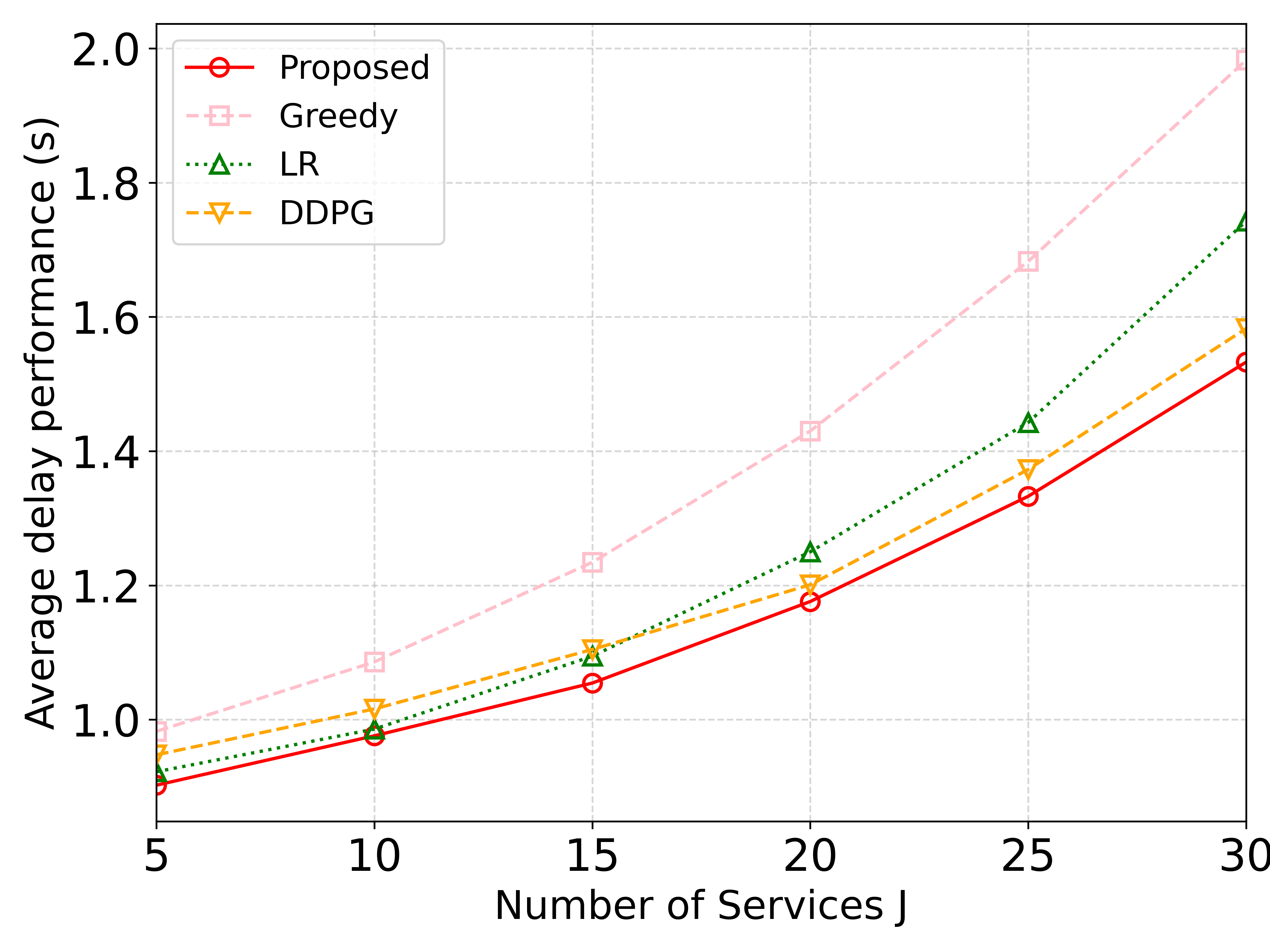}%
}
\vspace{-0.2cm}
\caption{Performance comparison of algorithms under (a) different numbers of cells and (b) different numbers of services.}
\label{fig5}
\end{figure}
As shown in Fig. \ref{fig5}(a), the average processing delay under all schemes gradually decreases as the number of cells increases, which benefits from the expanded cooperation scope as more cells are involved. Moreover, the performance gap between the baseline schemes and the proposed scheme widens as the number of cells increases. As illustrated in Fig. \ref{fig5}(b), the average processing delay of all schemes increases with the number of services, mainly due to the limited storage capacity of ESs. Notably, unlike the trend in Fig. \ref{fig5}(a), only the Popularity-based and Greedy schemes show an increasing performance gap from the proposed scheme as the number of services grows. Furthermore, when the number of services is relatively small ($J < 15$), the Popularity-based scheme outperforms the DDPG scheme, as the storage capacity of ESs is no longer the performance bottleneck in this case. 
\vspace{-0.2cm}
\begin{figure}[h]
\centering
\subfloat[]{\includegraphics[width=0.24\textwidth]{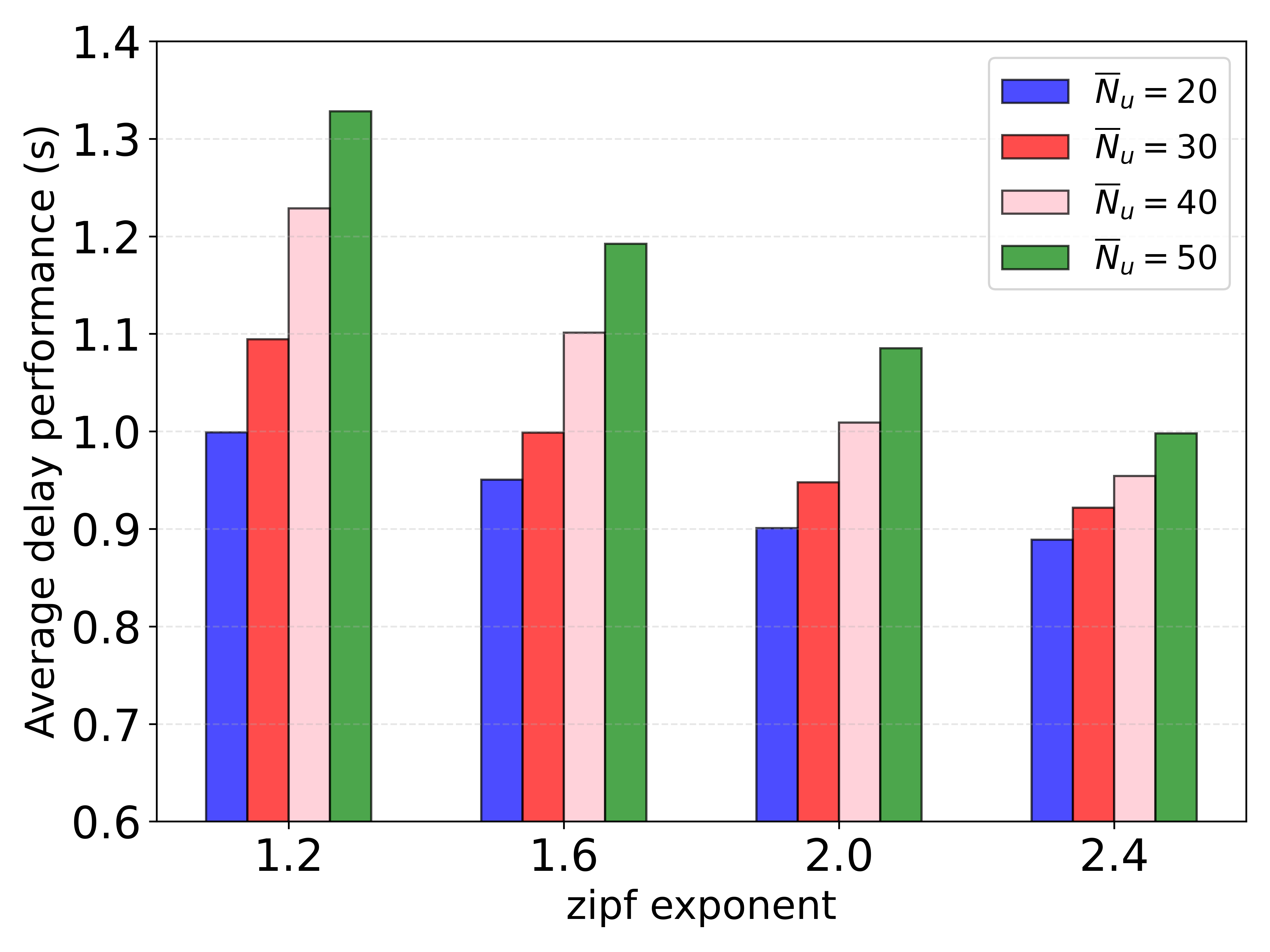}%
}
\subfloat[]{\includegraphics[width=0.24\textwidth]{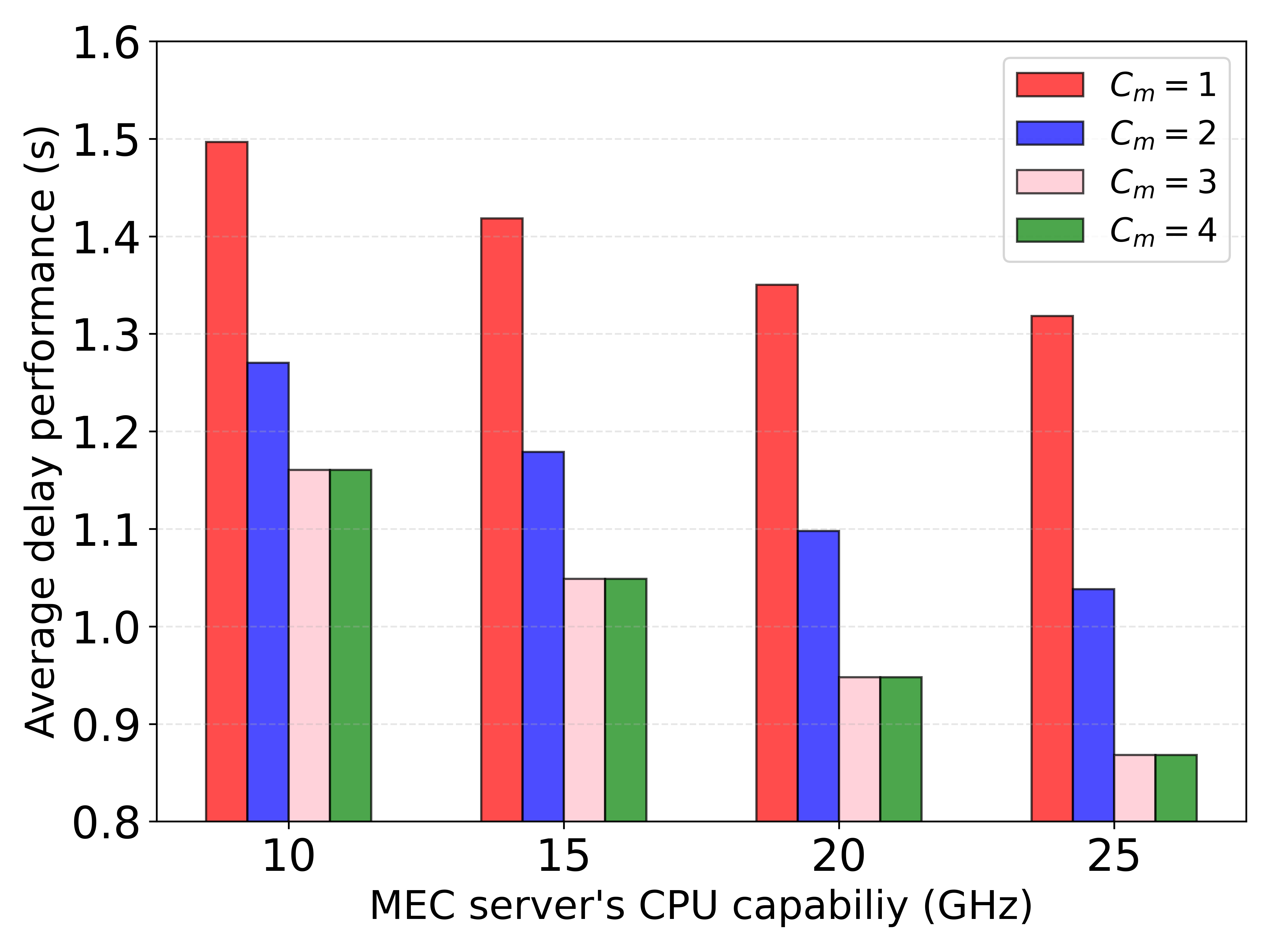}%
}
\vspace{-0.2cm}
\caption{Impact of System Parameters on Algorithm Performance.}
\label{fig6}
\end{figure}

As shown in Fig. \ref{fig6}(a), the average processing delay achieved by the proposed algorithm gradually decreases with an increasing Zipf exponent. This occurs because a higher Zipf exponent concentrates service requests, thereby reducing the influence of limited ES storage capacity on optimal service deployment. Moreover, the average delay increases with the average number of users $\overline{N}_u$ in each cell. As illustrated in Fig. \ref{fig6}(b), increasing the CPU or storage capacity of ESs reduces the average delay. Notably, when the storage capacity of ESs is sufficiently large ($C_m > 3$), further increases do not lead to additional reductions in the average delay. This is because storage capacity is no longer the limiting factor; instead, the computational capability of ESs becomes the bottleneck.

\section{Conclusion}\label{sec:conclusion}
This paper investigates the joint service deployment and task scheduling problem in collaborative edge networks, with the objective of minimizing long-term expected task processing delay. To address this, we propose a two-timescale optimization framework. The proposed approach integrates convex optimization with multi-agent DRL, enabling coordinated optimization across temporal and spatial dimensions through online alternating optimization at two distinct timescales. Future work will extend the current framework by incorporating resource allocation and optimizing additional objectives such as energy efficiency and overall resource utilization.

\section*{Acknowledgment}
This work was supported by the National Science Foundation of China under Grant 62271062 and the National Project under Grant 2024-JCJQ-ZD-050-02.

\bibliographystyle{IEEEtran}
\bibliography{ref_new}

\end{document}

%% file: glossary.tex
\makeglossaries
\newacronym{6g}{6G}{Sixth Generation}
\newacronym{awgn}{AWGN}{additive white Gaussian noise}
\newacronym{bcd}{BCD}{block coordinate descent}
\newacronym{bri}{BRI}{biregular irreducible}
\newacronym{cp}{CP}{control plane}
\newacronym{csi}{CSI}{channel state information}
\newacronym{csit}{CSIT}{channel state information at transmitter}
\newacronym{dft-s-ofdm}{DFT-s-OFDM}{Discrete Fourier Transform-spread-OFDM}
\newacronym{fbl}{FBL}{finite blocklength}
\newacronym{gan}{GAN}{generative adversarial network}
\newacronym{ibl}{IBL}{infinite blocklength}
\newacronym{lfp}{LFP}{leakage-failure probability}
\newacronym{mcs}{MCS}{modulation and coding scheme}
\newacronym{mimo}{MIMO}{multi-input multi-output}
\newacronym{mm}{MM}{Minorize-Maximization}
\newacronym{noma}{NOMA}{non-orthogonal multi-access}
\newacronym{nom}{NOM}{non-orthogonal multiplexing}
\newacronym{ofdm}{OFDM}{orthogonal frequency-division multiplexing}
\newacronym{ofdma}{OFDMA}{orthogonal frequency-division multiple access}
\newacronym{oma}{OMA}{orthogonal multiple access}
\newacronym{papr}{PAPR}{Peak-to-Average Power Ratio}
\newacronym{pdf}{PDF}{probability density function}
\newacronym{per}{PER}{packet error rate}
\newacronym{phy}{PHY}{physical}
\newacronym{pld}{PLD}{physical layer deception}
\newacronym{pls}{PLS}{physical layer security}
\newacronym{prb}{PRB}{physical resource block}
\newacronym{sic}{SIC}{successive interference cancellation}
\newacronym{sinr}{SINR}{signal-to-interference-and-noise ratio}
\newacronym{snr}{SNR}{signal-to-noise ratio}
\newacronym{tdma}{TDMA}{time-division multiple access}
\newacronym{up}{UP}{user plane}
\newacronym{urllc}{URLLC}{ultra-reliable low-latency communication}